\documentclass[sigconf]{acmart}
\AtBeginDocument{%
  \providecommand\BibTeX{{%
    \normalfont B\kern-0.5em{\scshape i\kern-0.25em b}\kern-0.8em\TeX}}}

\copyrightyear{2022}
\acmYear{2022}
\setcopyright{acmcopyright}\acmConference[WWW '22]{Proceedings of the ACM Web
Conference 2022}{April 25--29, 2022}{Virtual Event, Lyon, France}
\acmBooktitle{Proceedings of the ACM Web Conference 2022 (WWW '22), April
25--29, 2022, Virtual Event, Lyon, France}
\acmPrice{15.00}
\acmDOI{10.1145/3485447.3512215}
\acmISBN{978-1-4503-9096-5/22/04}
\usepackage{hyperref}       % hyperlinks
\usepackage{url}            % simple URL typesetting
\usepackage{booktabs}       % professional-quality tables
\usepackage{amsfonts}       % blackboard math symbols
\usepackage{nicefrac}       % compact symbols for 1/2, etc.
\usepackage{microtype}      % microtypography
\usepackage{xcolor}         % colors
\usepackage{algorithm}
\usepackage{algorithmic}
\usepackage{graphicx}
\usepackage{mathtools}
\usepackage{amsmath}
\usepackage{dsfont}
\usepackage{multirow}
\usepackage{enumitem}
\usepackage{soul}
\newcommand{\mymethod}{\textsf{ACORN}}

\begin{document}

\title{Socialbots on Fire: Modeling Adversarial Behaviors of Socialbots via Multi-Agent Hierarchical Reinforcement Learning}

% \author{Annonymous Author(s) - Submission \#120}
\author{Thai Le}
\email{tql3@psu.edu}
\affiliation{%
  \institution{Penn State University, USA}
}

\author{Long Tran-Thanh}
\email{long.tran-thanh@warwick.ac.uk}
\affiliation{%
  \institution{University of Warwick, UK}
%   \country{UK}
}

\author{Dongwon Lee}
\email{dongwon@psu.edu}
\affiliation{%
  \institution{Penn State University, USA}
}


\begin{abstract}
Socialbots are software-driven user accounts on social platforms, acting autonomously (mimicking human behavior), with the aims to influence the opinions of other users or spread targeted misinformation for particular goals. As socialbots  undermine the ecosystem of social platforms, they are often considered harmful. As such,
there have been several computational efforts to auto-detect the socialbots. However, to our best knowledge, the {\em adversarial} nature of these socialbots has not yet been studied. This begs a question ``can  adversaries, controlling socialbots, exploit AI techniques to their advantage?" To this question, we successfully demonstrate that indeed it is possible for adversaries to exploit computational learning mechanism such as reinforcement learning (RL) to maximize the influence of socialbots while avoiding being detected. We first formulate the adversarial socialbot learning as a cooperative game between two functional hierarchical RL agents. While one agent curates a sequence of activities that can avoid the detection, the other agent aims to maximize network influence by selectively connecting with right users. Our proposed policy networks train with a vast amount of synthetic graphs and generalize better than baselines on {\em unseen real-life graphs} both in terms of maximizing network influence (up to +18\%) and sustainable stealthiness (up to +40\% undetectability) under a strong bot detector (with 90\% detection accuracy). During inference, the complexity of our approach scales linearly, independent of a network's structure and the virality of news. This makes our approach a  practical adversarial attack when deployed in a real-life setting.
\end{abstract}

\begin{CCSXML}
<ccs2012>
   <concept>
       <concept_id>10010147.10010257.10010258.10010261.10010272</concept_id>
       <concept_desc>Computing methodologies~Sequential decision making</concept_desc>
       <concept_significance>500</concept_significance>
       </concept>
 </ccs2012>
\end{CCSXML}
% \ccsdesc[500]{Security and privacy~Social network security and privacy}
\ccsdesc[500]{Computing methodologies~Sequential decision making}

\keywords{socialbot, social bot, adversarial, reinforcement learning}

\maketitle

\section{Introduction}

Socialbots refer to automated user accounts on social platforms that attempt to behave like real human accounts, often controlled by either automatic software, human, or a combination of both--i.e., cyborgs~\cite{cresci2020decade}. Different from traditional spambots, which may not have proper profiles or can be easily differentiated from regular accounts, socialbots often mimic the profiles and behaviors of real-life users by using a stolen profile picture or biography, building legitimate followships, replying to others, etc.~\cite{cresci2020decade}. Socialbots are often blamed for spreading divisive messages--e.g., hate speech, disinformation, and other low-credibility contents that have been shown to widen political divides and distrust among both online and offline communities~\cite{cresci2020decade,hindman2018disinformation,le2020malcom}. To mitigate such harmful proliferation of socialbots, therefore, there has been extensive research, most of which focus on how to effectively detect them~\cite{botwpi,dong2018feature,yang2020scalable}. However, these works usually follow the \textit{cat-and-mouse} game where they \textit{passively} wait for socialbot evasion to happen before they can react and develop a suitable detector~\cite{cresci2021coming}. Instead of following such a reactive scheme, however, proactively modeling socialbots and their adversarial behaviors on social platforms can better advance the next bot detection research.

In particular, we pose a question ``\ul{Can socialbots exploit computational learning mechanism such as reinforcement learning to their advantage?}"
To our best knowledge, {\em adversarial} nature of socialbots 
%and how to computationally model them 
has not yet been fully explored and studied. 
%That is, we pose a question ``\ul{Can socialbots exploit computational learning mechanism such as reinforcement learning to their advantage?}" 
However, it is plausible that adversaries who own a farm of socialbots operate their socialbots according to certain strategies (or algorithms). Therefore, proactively simulating such a computational learning mechanism and understanding adversarial aspect of socialbots better would greatly benefit future research on socialbot detection.

In general, a socialbot has two main objectives that are adversarial in nature: (i) to facilitate mass propaganda propagation through social networks and (ii) to evade and survive under socialbot detectors. The first goal can be modeled as an NP-Hard \textit{influence maximization} (IM) problem~\cite{kempe2003maximizing} where the bot needs to build up its network of followers--i.e., seed users, overtime such that \textit{any new messages} propagated from the bot through these users can effectively spread out and influence many other people. \textit{Simultaneously}, it also needs to systematically constrain its online behaviors such that it will not easily expose itself to socialbot detectors. Although the IM problem has been widely studied by several works~\cite{kempe2003maximizing,chen2009efficient,kingi2020numerical,kamarthi2019influence}, they only focus on maximizing the network influence given a \textit{fixed and static budget} \# of seed nodes (that is relatively \textit{small}) and they assume that every node is \textit{equally} acquirable. However, these assumptions are not practical in our context. Not only a socialbot needs to continuously select the next best seed node or follower over a long temporal horizon--i.e., potentially large budget of seed nodes, it also needs to consider that gaining the followship from a very influential actor--e.g., Elon Musk, is practically much more challenging than from a normal user. At the same time, a socialbot that optimizes its network of followers must also refrain from making suspicious behaviors--e.g., constantly following others, that can trigger the attention of bot detectors. Thus, learning how to navigate a socialbot is a very practical yet challenging task with two intertwined goals that cannot be separately optimized. 
%\lee{this paragraph is not super clear and fitting to intro}
Toward this challenge, in this paper, we formulate the \textbf{Adversarial Socialbot Learning (ASL)} problem and design a multi-agent {\em hierarchical reinforcement learning (HRL)} framework to tackle it. 

3Our main contributions are as follows. 

\begin{itemize}[leftmargin=\dimexpr\parindent+0.1\labelwidth\relax]
\item First, we formulate a novel ASL problem as an optimization problem with constraints. 
\item Second, we propose a solution to the ASL problem by framing it as a cooperative game of two HRL agents that represent two distinctive functions of a socialbot, namely (i) selecting the next best activity--e.g., tweet, retweet, reply, mention, and (ii) selecting the next best follower. We carefully design the RL agents and exploit unsupervised graph representation learning to minimize the potential computational cost resulted from a long time horizon and a large graph structure. 
\item Third, we demonstrate that such RL agents can \textit{learn from synthetic graphs yet generalize well on \textbf{real unseen graphs}}. Specifically, our experiments on a real-life dataset show that the learned socialbot outperforms baselines in terms of influence maximization while sustaining its longevity by continuously evading a strong black-box socialbot detector of 90\% detection accuracy. During inference, in addition, the complexity of our approach scales linearly and is independent of a network's structure and the virality of news. 
\item Four, we release an environment under the Open AI's \textit{gym}~\cite{gym} library. This enables researchers to simulate various adversarial behaviors of socialbots and develop novel bot detectors in a \textit{proactive} manner.
\end{itemize}

% \vspace{-0.1in}
\section{Related Work}

\subsection{Socialbots Detection}
The majority of previous computational works on socialbots within the last decade \cite{botwpi,yang2020scalable,dong2018feature,rodriguez2020one,sayyadiharikandeh2020detection,cai2017behavior,wu2021novel} primarily focus on developing computer models to effectively detect bots on social networks~\cite{cresci2020decade,cresci2021coming}. These models are usually trained on a ground truth dataset using supervised learning algorithms--e.g., Random Forest, Decision Tree, SVM, to classify an individual social media account into a binary label--i.e., bot or legitimate~\cite{cresci2020decade}. Moreover, these learning algorithms usually depend on either a set of statistical engineered predictive features such as the number of followers, tweeting frequency, etc.~\cite{yang2020scalable,sayyadiharikandeh2020detection,cresci2017paradigm}, or a deep learning network where the features are automatically learned from unstructured data such as an account's description text. Even though there are many possible features that can be used to detect socialbots, statistical features that can be directly extracted from user metadata provided by official APIs--e.g., Twitter API, are more practical due to their favorable computational speed in practice~\cite{yang2020scalable}. In fact, many of the features that are utilized by the popular socialbot detection API \textit{botometer} fall into this category. Moreover, we later also show that using simple statistical features derived from user metadata can help train a socialbot detector with around 90\% prediction accuracy on a hold-out test set (Sec. \ref{sec:environment}). Regardless of how a socialbot detector extracts its predictive features, they are mainly designed following a \textit{reactive} schema where they learn how to detect socialbots after they appear (thus a training dataset can be collected).

\subsection{Adversarial Socialbot Learning}

While previous works help us to understand better the \textit{detection} aspect of socialbots, the \textit{learning} aspect of them has not been widely studied~\cite{cresci2021coming}. Distinguished from learning how to detect socialbots using a stationary snapshot of their features, ASL computationally models the adversarial learning behaviors of socialbots over time. To the best of our knowledge, relevant works on this task are limited to \cite{cresci2019better}. This work adopts an evolution optimization algorithm to find different adversarial permutations from a \textit{fixed} socialbot' encoded activity sequence--e.g., ``tweet$\rightarrow$tweet$\rightarrow$retweet$\rightarrow$reply,...", and examine if such permutations can help improve the detection accuracy of a bot detector. However, such permutations, even though adversarial in nature, are just static snapshots of a socialbot and do not tell a whole story on how the bot evolves. In other words, we are still lacking a general computation framework that models the temporal dynamics of socialbots and their adversarial behaviors. Therefore, this paper aims to formally formulate their behaviors as a Markov Decision Process (MDP)~\cite{howard1960dynamic} and designs an RL framework to train socialbots that can optimize their adversarial goals on real-life networks.

We investigate two adversarial objectives of a socialbot: influencing people while evading socialbot detection. While the first one can be modeled as an IM task on graph networks, traditional IM algorithms--e.g.,\cite{kempe2003maximizing,chen2009efficient,kingi2020numerical}, assume that the number of seed nodes is relatively small and all nodes are equally acquirable, all of which are not applicable in the socialbot context as previously described. There have been also a few works--e.g., \cite{li2019disco,tian2020deep}, that utilizes RL to IM task. Yet their scope is still limited to a single constraint on the budget number of seeds. Influence maximization under a temporal constraint--i.e., not to be detected lead to early termination in this case, is a non-trivial problem.

\section{Problem Formulation}

\subsection{Social Network Environment}\label{sec:environment}
\textbf{Network Representation and Influence Diffusion Model}
A social network includes users, their interactions and how they influence each other. We model this network as a directed graph $G{=}(V,E)$. An edge between two users $u,v{\in} V$, denoted as $(u,v){\in} E$, means $u$ can have influence on $v$. $(u,v)$ also illustrates a piece of news can spread from $u$ to $v$--i.e, $v$ follows $u$ (thus $u$ influences $v$).

As there is no influence model that can perfectly reflect real-world behaviors, to model the influence flow through $G$, we adopt \textit{Independence Cascade Model (ICM)}~\cite{goldenberg2001talk,goldenberg2001using}, which is the most commonly used in the context of a social network~\cite{kimura2006tractable,jendoubi2017two,li2017survey}. ICM was originally proposed to model the “word-of-mouth” behaviors, which resemble the information sharing phenomena online well. In ICM, a node is either active or inactive. Once a node $u$ is activated, it has a single opportunity to activate or influence its inactive neighbors $\mathcal{N}(u)$ with an \textit{uniform} \textit{activation probability} $p$. At first, every node is inactive except a set of seed nodes $\mathcal{S}$. After that, as the environment rolls out throughout a sequence of discrete timesteps, the influence will propagate from $\mathcal{S}$ through the network by activating different nodes in $G$ following $E$ and $p$. The process ends when there is no additional activated nodes being activated~\cite{kamarthi2019influence,li2021claim}. Hence $p$ is also the virality of news--i.e., how fast a piece of news can travel through $G$. We then use $G{=}(V,E,p)$ to denote the social network $G$. 

Let denote by $\pmb{\sigma}(\mathcal{S}, G)$ the \textit{spread function} that measures how many nodes in $G$ a piece of information--e.g., fake news, can spread from $\mathcal{S}$ via the ICM model. Given a fixed network structure $(V,E)$ and the news virality $p$, different $\mathcal{S}$ will result in different values of $\pmb{\sigma}(\mathcal{S},G)$. Hence, selecting a good $\mathcal{S}$ is decisive in optimizing the spread of influence on $G$. However, choosing $\mathcal{S}$ to maximize $\pmb{\sigma}(\mathcal{S}, G)$ has already been proven to be an NP-Hard problem~\cite{kempe2003maximizing}.
\\

\noindent \textbf{Socialbots.} A socialbot is then a vertex in $G$ that attempts to mimic human behaviors for various aims--e.g., spreading propaganda or low-credible contents through $G$, ~\cite{shao2018spread,cresci2020decade,subrahmanian2016darpa}. It carries out a sequence of activities $\mathcal{A}$ to \textit{simultaneously} achieve two main objectives:

\begin{enumerate}[leftmargin=\dimexpr\parindent+1.0\labelwidth\relax, label=\textbf{Obj. \arabic*:}]
    \item \textit{Optimizing its influence over $G$ by selectively collecting good seed nodes--i.e., {followers}, $\mathcal{S}{\in} V$, over time}
    \item \textit{Evading bots detectors--i.e., not to be detected and removed}
\end{enumerate}

These two goals are often in tension in that improving Obj 1 typically hurts Obj 2 and vice versa. That is while having a good network of followers $\mathcal{S}$ enables a socialbot to spread disinformation to a large number of users at any time, having a high undetectability helps it to sustain this advantage over a long period. As socialbots are usually deployed in groups, and later coming socialbots can also easily inherit a previously established network of followers $\mathcal{S}$ of a current one. If a bot is detected and removed from $G$, not only it can lose its followers $\mathcal{S}$ and expose itself to be used to develop stronger detectors, it can also risk revealing the identity of other bots--e.g., by way of guilt-by-association~\cite{wang2017gang}. This makes the sustainability achieved through \textit{Obj 2} distinguishably important from previous literature--e.g., ~\cite{kamarthi2019influence,kempe2003maximizing,wen2016online}, where the optimization of $\mathcal{S}$ plays a more central role.
\\

\renewcommand{\tabcolsep}{6pt}
\begin{table}[tb]
    \centering
    \small
    \caption{Predictive features of the socialbot detector $\mathcal{F}$.}
    \begin{tabular}{lr}
        \toprule
        \multicolumn{1}{c}{{Feature}} & \multicolumn{1}{c}{{Description}} \\
        \cmidrule(lr){1-2}
        \#tweets & \# of tweets posted by the user \\
        \#replies & \# of replies posted by the user \\
        \#retweets & \# of retweets posted by the user \\
        \cmidrule(lr){1-2}
        \#avg.tweets & average \# tweets posted per timestep \\
        \#avg.replies & average \# replies posted per timestep \\
        \#avg.retweets & average \# retweets posted per timestep \\
        \cmidrule(lr){1-2}
        \#retweet.ratio  & \#retweets/\#tweets \\
        \#replies.ratio  & \#replies/\#tweets \\
        \#retweet.replies.ratio  & \#retweets/\#replies \\
        % \cmidrule(lr){1-2}
        \#mentions.ratio & \# unique mentions posted per tweet \\
        % \#followship.ratio & \# followers/\# following \\
        % \cmidrule(lr){1-2}
        % DNA & unique 5-gram features from DNA signature \\
        \bottomrule
    \end{tabular}
    \vspace{-10pt}
    \label{tab:features}
\end{table}

\noindent \textbf{Relationship between $\mathcal{A}$ and $\mathcal{S}$.} $\mathcal{A}$ denotes the activity sequence--i.e., the DNA of the bot~\cite{cresci2017social}. $\mathcal{A}$ includes four possible types of actions to be made at every timestep $t$, namely \textit{tweet, retweet, reply or mention}, and only the \textit{last three} of which can directly interact with others to expand $\mathcal{S}$. Despite these actions are in the \textit{Twitter} context, other platforms also provide similar functions--e.g., tweet->post, retweet->share, reply->comment, mention->tag on Facebook. In practice, \textit{not} every node requires an equal effort to convert to a follower. For example, a bot needs to accumulate its reputability over time and interact more frequently to have an influencer--e.g., Elon Musk, rather than a normal user to become its follower. Since a real model underlining such observation is unknown, we model it using a simple heuristic:
\begin{equation}
\begin{aligned}
    g_{Q}(u, t) &= \mathrm{max}(1, Q{f}(u, t)) \quad \mathrm{where} \\
    \quad {f}(u, t) &\sim \mathrm{Bernoulli}(1 - \frac{1+|\mathcal{S}_t|}{1+|\mathcal{N}(u)|}), 
\label{eqn:dependency}
\end{aligned}
\end{equation}

where $g_{Q}(u,t)$ with hyper-parameter $Q{\geq}1$, is the \textbf{number of times} the socialbot is \textit{required} by the environment to continuously interact with an influencer $u$--i.e., high $\mathcal{N}(u)$, for it to become a follower at $t$. Intuitively, a bot with a good reputation overtime--i.e., a high number of followers at the timestep $t$--i.e., $|\mathcal{S}_t|$, can influence others to follow itself more effortlessly than a newly created bot. Overall, $\mathcal{A}$ encodes when and what type of interaction--i.e., \textit{retweet, reply} or \textit{mention}, to use to acquire a new follower $s{\in}\mathcal{S}$, $s$ then decides the \textit{frequency} of such interaction in $\mathcal{A}$. Thus, $\mathcal{A}$ and $\mathcal{S}$ is temporally co-dependent.
\\

\noindent \textbf{Socialbot Detection Model.}
Bot detectors are responsible for detecting and removing socialbots from $G$. Let $\mathcal{F}(\mathcal{A}_t){\in}\{0,1\}$ denote a model that predicts whether or not an account is a socialbot based on its activity sequence \textit{up to} the timestep $t$ ($\mathcal{A}_t$). This sequence of ordered activity is then usually represented as an \textit{unordered list} of statistical features such as number of replies, tweets per day, by socialbot detectors~\cite{botwpi,dong2018feature,yang2020scalable}. In this paper, $\mathcal{F}$ extracts and adopts several features (Table \ref{tab:features}) from previous works for detection. Most of the features are utilized by the popular bot detection API \textit{Botometer}~\cite{davis2016botornot}. We train $\mathcal{F}$ using the \textit{Random Forest}~\cite{svetnik2003random} algorithm with supervised learning on a \textit{publicly available dataset}~\cite{yang2019arming,mazza2019rtbust}~\footnote{\url{https://botometer.osome.iu.edu/bot-repository/}} of nearly 15K Twitter accounts, half of which is labelled as socialbots. This dataset is \textit{not} exposed to the socialbots. Here we also assume that $\mathcal{F}(\cdot)$ is a \textit{black-box model}--i.e., we do not have access to its parameters. $\mathcal{F}$ achieves nearly 90\% in F1 score on an unseen test set following the standard 5-fold cross validation (train and test with 80\%/20\% data). Since $\mathcal{A}$ and $\mathcal{S}$ are co-dependent, we can easily see that $\mathcal{S}$ also has effects on the detectability of a socialbot. 
Note that to focus on the study of the adversarial aspect of socialbots, we had to resort to a certain combination of account features and the socialbot detection model. 90\% in F1 score is also in line with SOTA detectors on a similar set of features~\cite{botwpi}.
%\lee{needs to explain why our choice of  features/random-forest/twitter-data was necessary, and does not affect our later findings negatively}

 \begin{figure*}[t!]
  \centering
  \includegraphics[width=0.70\textwidth]{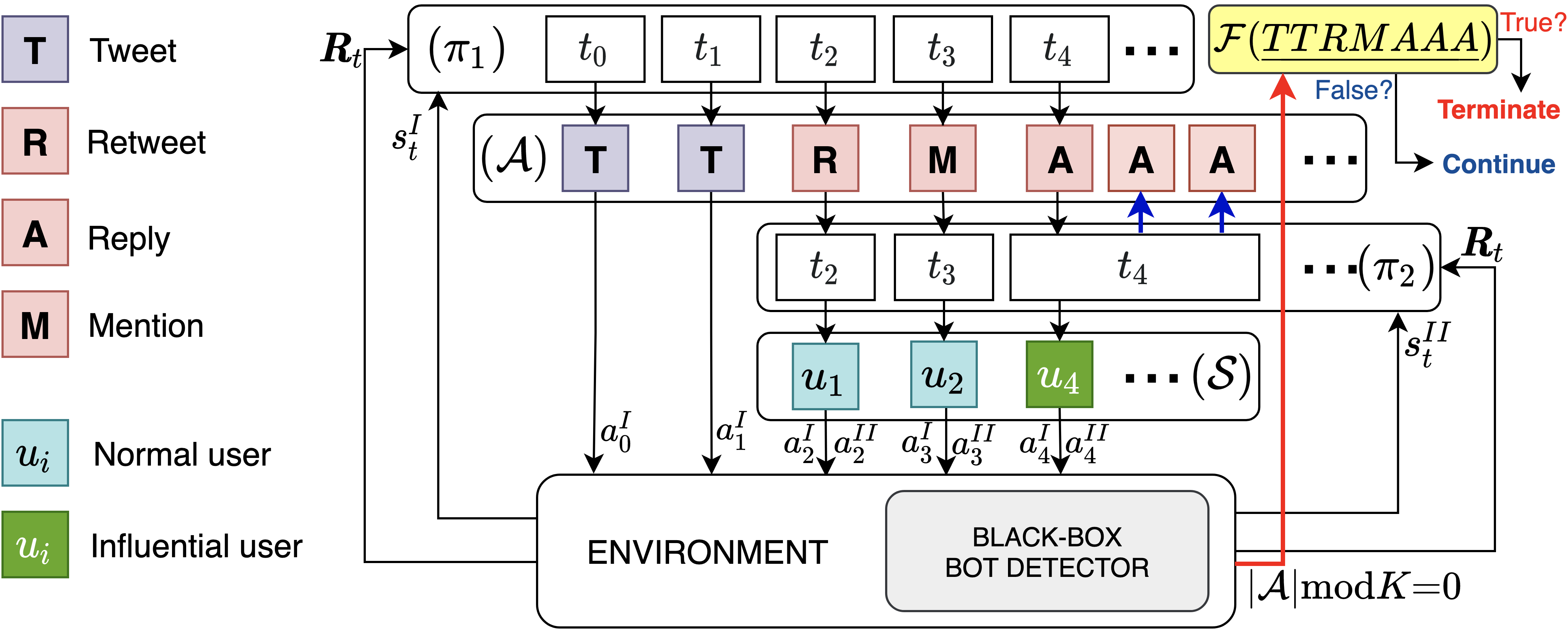}
  \caption{An example of \textsc{\mymethod} HRL framework. As the environment rolls out, \textsc{AgentI} ($\pi_1$) decides which type of activity (T, R, A or M) to perform. Whenever an interactive action (R, A, M) is selected, \textsc{AgentII} ($\pi_2$) then selects a new follower. Since the selected user $u_4$ at $t_4$ is an influencer, $\pi_2$ needs perform not once but $Q{=}3$ times of action ``A" to acquire $u_4$ (\textcolor{blue}{blue} arrow). Whenever $|\mathcal{A}|$ reaches an interval of $K{=}7$, the bot detector $\mathcal{F}(\mathcal{A}_t)$ is triggered (\textcolor{red}{red} arrow).
  } 
  \label{fig:framework}
  \vspace{-10pt}
\end{figure*}

\subsection{The ASL Problem and Objective Function}

From the above analysis, this paper proposes to study the Adversarial Socialbot Learning (ASL) problem to achieve both \textit{Obj 1} and \textit{Obj 2}. In other words, we aim to solve the following problem.

\noindent
\begin{center}
\fbox{\parbox[t]{1.0\linewidth}{
\textbf{\textsc{Problem}}: Adversarial Socialbot Learning (ASL) aims to develop an automatic socialbot that can exercise adversarial behaviors against a black-box bot detector $\mathcal{F}$ while at the same time maximizing its influence on $G$ through a set of selective followers $\mathcal{S}$.
\label{problem}
}}
\end{center}

Specifically, we formulate this task as an optimization problem with the objective function as follows.

% \begin{subequations}
% \begin{align}
%     &\max_{\mathcal{S}_t, \mathcal{A}_t} \quad \pmb{R^*} = \pmb{\sigma}(\mathcal{S}_{T^*}, G)(1 + T^*) \quad \textrm{subject to}\\ 
%     &T^* = \min_{T^*}\;\Big[\mathcal{F}(\mathcal{A}_{T^*}) = 1 \land \mathcal{F}(\mathcal{A}_t) = 0 \Big] \\
%     &\quad \quad \quad {\Large\forall}\; 1{<}t{<}T^*, |\mathcal{A}_t|{\bmod} K{=}|\mathcal{A}_{T^*}|{\bmod}K{=}0 \\
%     &g_Q(u, t) = max(1,Q{f}(u, t)) \quad {\Large\forall}\; 1{<}t{<}T^*
% \end{align}
% \label{eqn:objective}
% \end{subequations}

\begin{center}
\fbox{\parbox[t]{1.0\linewidth}{
\textbf{\textsc{Objective Function}}: Given a black-box bot detection model $\mathcal{F}$ and a social network environment what is characterized by $G{=}(V,E,p)$, $K$, $Q$, we want to optimize the objective function: 
\begin{subequations}
\begin{align}
    &\max_{\mathcal{S}_t, \mathcal{A}_t} \quad \pmb{R^*} = \pmb{\sigma}(\mathcal{S}_{T^*}, G)(1 + T^*) \quad \textrm{subject to}\\ 
    &T^* = \min_{T^*}\;\Big[\mathcal{F}(\mathcal{A}_{T^*}) = 1 \land \mathcal{F}(\mathcal{A}_t) = 0 \Big] \\
    &\qquad \quad \quad {\Large\forall}\; 1{<}t{<}T^*, |\mathcal{A}_t|{\bmod} K{=}|\mathcal{A}_{T^*}|{\bmod}K{=}0 \\
    &g_Q(u, t) = max(1,Q{f}(u, t)) \quad {\Large\forall}\; 1{<}t{<}T^*
\end{align}
\vspace{-10pt}
\label{eqn:objective}
\end{subequations}}}
\end{center}
\noindent Socialbot detector $\mathcal{F}$ can run prediction on the socialbot every time it performs a new activity. However, $\mathcal{A}_u$ and $|V|$ can potentially be very large. Thus, we assume that $\mathcal{F}$ only runs detection every time $K$ new activities is added to $\mathcal{A}$ (Eqn. \ref{eqn:objective}b). This makes $T^*$ the earliest interval timestep at which a socialbot is detected and removed by $\mathcal{F}$ (Eqn. \ref{eqn:objective}b,c). Since $\pmb{R^*}$ is \textit{monotonically increasing} on both $V{\geq}\pmb{\sigma}(\mathcal{S}_b, G){\geq}0$ and $T^*{\geq}1$, to maximize $\pmb{R^*}$, a socialbot cannot focus \textit{only} either on \textit{Obj 1} or \textit{Obj 2}. In other words, Eqn. (\ref{eqn:objective}d) encourages the socialbot to simultaneously optimize both objectives.
% \ltt{Perhaps link each explanation to the corresponding line in the formulation? It took me a while to see what the maths parts mean.}
% Since, $\pmb{\sigma}(\mathcal{S}_b, G)\leq |V|$, a bot can optimize Eq. (\ref{eqn:objective}) by maximizing only $T^*$ with $|S|=0$. Therefore, $\beta_1$ and $\beta_2$ are introduced to help control the priority between \textit{Obj 1} and \textit{Obj 2} over time. 

\section{The Proposed Method: \textsc{\mymethod}}
% In this section, we model the ASL task as a contextual Markov Decision Process (MDP)~\cite{hallak2015contextual,modi2020no}. 

\subsection{Markov Decision Process Formulation}\label{sec:MDP}
 The ASL problem can be formulated as an MDP process which consists of a state set $S$, an action set $A$, a transition function $\mathcal{P}$, a reward function $R$, a discount factor $\gamma\in [0,1]$ and the horizon $T$. 
%  The state $S$ includes the network's structure $G{=}(V,E,p)$, the current followers $\mathcal{S}$ and action sequence $\mathcal{A}$ at each timestep $t$. The action set $A$ includes 4 possible activities (tweet, retweet, reply and mention) that construct $\mathcal{A}$ and $|V|$ possible options for followers. A standard RL algorithm then aims to maximize the expected discounted return $\eta(\pi_\theta)$, 
% %  $\eta(\pi_\theta){=}\mathbb{E}_\tau([\sum_{t=0}^{T^*} \gamma^t r(s_t, a_t)])$
%  where $T^*{\leq}T$, $\tau{=}(s_0, a_0,...)$ stores the whole trajectory, $a_t \sim \pi_{\theta}(a_t|s_t)$ is a policy parameterized by $\theta$ that specifies an action distribution for each state $s_t$ and $s_{t+1} \sim \mathcal{P}(s_{t+1}|s_t, a_t)$. The state $s_0$ is initialized with $G{=}(V,E,p),\mathcal{S}_0{=}\mathcal{A}_0{=}\emptyset$.
 Since the space requirement for $A$ can be very large--i.e., $4|V|$ for 4 possible activities and $|V|$ possible seed nodes, especially on a large network, this can make the task much more challenging to optimize due to potential sparse reward problem. To overcome this, we transformed this into a HRL framework of two functional agents, \textsc{AgentI} and \textsc{AgentII}, with a \textit{global reward} (Figure \ref{fig:framework}). We call this \textsc{\mymethod}
 (\ul{A}dversarial so\ul{C}ialb\ul{O}ts lea\ul{R}ni\ul{N}g)
 framework. While \textsc{AgentI} is
 responsible for deciding which \textit{type} of activity among \{tweet, retweet, reply, mention\} to perform at each timestep $t$, \textsc{AgentII} is mainly responsible for $\mathcal{S}$--i.e., to select which follower to accumulate, \textit{only when} \textsc{AgentI} chooses to do so--i.e., retweet, reply, mention. This reduces the overall space of $A$ to only $|V|{+}4$. Since $\mathcal{A}$ and $\mathcal{S}$ are co-dependent (Sec. \ref{sec:environment}), the two agents need to continuously cooperate to optimize both influence maximization and undetectability. It is noted that the Markov assumption behind this MDP is not violated because both influence function $\pmb{\sigma}(\cdot)$ and detection probability $\mathcal{F}$ at time $t$ only depends on statistical snapshot of the two agents at $t{-}1$. This HRL task is then described in detail as follows.
 \\

\noindent  \textbf{State.} Following ~\cite{florensa2017stochastic,li2019hierarchical,konidaris2007building}, we assume that the state space $S$ can be factorized into bot-specific $S_{\mathrm{DNA}}$ and network-specific $S_{\mathrm{ENV}}$, and $s^I{\in} S_{\mathrm{DNA}}, s^{II}{\in} S_{\mathrm{ENV}}$, where $s^I, s^{II}$ is the state space of \textsc{AgentI} and \textsc{AgentII}, respectively. Specifically, $s^I_t$ encodes (i) the number of followers $|S|$ of the bot and (ii) a snapshot of $\mathcal{A}_t$ at timestep $t$. While $s^I_t$ can directly store the actual $\mathcal{A}_t$ sequence, this potentially induces a computational and space overhead especially when $t$ becomes very large. Instead, we compact $\mathcal{A}_t$ into a fixed vector summarizing the frequency of each \textit{tweet}, \textit{retweet}, \textit{reply}, and \textit{mention} action up to $t$. This effectively limits the space complexity of $s^I{\in}\mathbb{R}^5$ to $\mathcal{O}(1)$. Similarly, $s^{II}_t\in \mathbb{R}^{4+|V|(k+1)}$ comprises of (i) $\mathrm{node2vec}(G)$~\cite{grover2016node2vec} which encodes the structure of $G$ to $|V|$ vectors of size $k$, (ii) a statistical snapshot of $\mathcal{A}_t$ and (iii) information regarding $\mathcal{S}_t$, encoded as:
 
 \begin{equation}
     (\mathds{1}(u{\notin}\mathcal{S}_t)\frac{1+|\mathcal{S}_t|}{1+|\mathcal{N}(u)|})_{u{=}0}^{|V|}\in \mathbb{R}^{|V|}
     \label{eqn:encode_s}
 \end{equation}
 
Previous works have often encoded the network structures (\cite{gogineni2020torsionnet,zhang2020learning}) via a parameterized Graph Neural Network (GCN)~\cite{kipf2017semi} as part of the policy network.
%which requires frequent parameter updates during training. 
As this approach requires frequent parameter updates during training, instead, we adopt \textit{node2vec}($G$) as an alternative unsupervised method which requires the calculation \textit{only once}. While $\mathcal{S}_t$ can be encoded as a one-hot vector $(\mathds{1}(u{\notin}\mathcal{S}_t))_{u=0}^{|V|}$, we enrich it by multiplying it with the binary ${f}(u,t)$ condition $\frac{1+|\mathcal{S}_t|}{1+|\mathcal{N}(u)|}$ (Sec. \ref{sec:environment}), which then results in Eq. (\ref{eqn:encode_s}). This enables \textsc{AgentII} to select nodes accordingly with the current reputation of the bot $|\mathcal{S}_t|$.
\\
% \lee{this sentence is unclear and grammatically odd}
% Moreover, \textit{node2vec} also enables us to directly control the exploration rate (through $p_{\mathrm{in}}$ and $p_{\mathrm{out}}$) to focus on learning the structure similarities among nodes rather than among communities in which they belong to. This is especially important in the ASL problem because $G$ usually includes several star-shape communities where information flows from a single influential user.
 
\noindent  \textbf{Action and Policy.} Similarly, we factor $A$ into two different action spaces $a^I, a^{II}$ for \textsc{AgentI} and \textsc{AgentII}, respectively. $a^I{\in} \mathbb{R}^4$, $a^{II}{\in} \mathbb{R}^{|V|}$ are both encoded as one-hot vectors, representing one of four available activities 
 %{tweet, retweet, reply and mention} 
 and one of potential followers, respectively. We then have two policies $\pi_1{=}(a^I|s^I)$, $\pi_2{=}(a^{II}|s^{II},a^{I})$ that control \textsc{AgentI} and \textsc{AgentII}, respectively. 
 \\
 
 \noindent \textbf{Reward.} Even though we can directly reward the RL agents with $\pmb{\sigma}(\mathcal{S}_t, G){\geq}1.0$ at every timestep $t{\leq}T^*$, this calculation will incur large computational cost, especially when $T^*$ becomes large. Instead, therefore, we design an accumulative reward function $R$ that consists of a \textit{step reward} and a \textit{delayed reward} to incentivize  RL agents to maximize $\pmb{R}^*$ (Eqn. \ref{eqn:objective}) as follows.
 \begin{equation}
 \begin{aligned}
     \pmb{R}_{\mathrm{step}}(t) &= \pmb{\sigma}(\mathcal{S}_t \setminus \mathcal{S}_{t-1}, G)\\ \pmb{R}_{\mathrm{delayed}}(T^*) &= \pmb{\sigma}(\mathcal{S}_{T^*}, G)
 \end{aligned}
 \end{equation}
where $T^*{\leq}T$ is the interval timestep at which the bot is detected and the episode is terminated. The step reward $\pmb{R}_{\mathrm{step}}$, which can be efficiently computed, is the \textit{marginal} gain on the network influence given a new follower selected at $t$. Using the step reward with a discount factor, $\gamma_{\mathrm{step}}{<}1.0$,  helps avoid the sparse reward problem and encourages good follower selection \textit{early} during an episode. Since $\pmb{R}_{\mathrm{step}}\geq1.0$, it also encourages the bot to survive against bot detection longer--i.e., to maximize $T^*$. In other words, as long as the socialbot survives--i.e., $T^*$ increases, in other to make new friendship, it will be able to influence more people. However, since $\pmb{\sigma}(\cdot)$ is \textit{subadditive}--i.e., $\pmb{\sigma}(\{u\},G){+}\pmb{\sigma}(\{v\}, G){\geq}\pmb{\sigma}(\{u,v\}, G) \;{\Large\forall}\; u,v{\in} V$, we then introduce the delayed reward $\pmb{R}_{\mathrm{delayed}}$ at the end of each episode with a discounted factor $\gamma_{\mathrm{delayed}}{<}1.0$ as a reward adjustments for each node selection step. 

% $\sum_{t=1}^{T^*}\pmb{R}_{\mathrm{step}}(t)\geq \pmb{R}_{\mathrm{delayed}}(T^*)$, 
% $\pmb{\sigma}(\{u\},G) + \pmb{\sigma}(\{v\}, G)\geq \pmb{\sigma}(\{u,v\}, G) \;{\Large\forall}\; u,v\in V$, 

 \begin{figure}[t!b]
  \centering
  \includegraphics[width=0.45\textwidth]{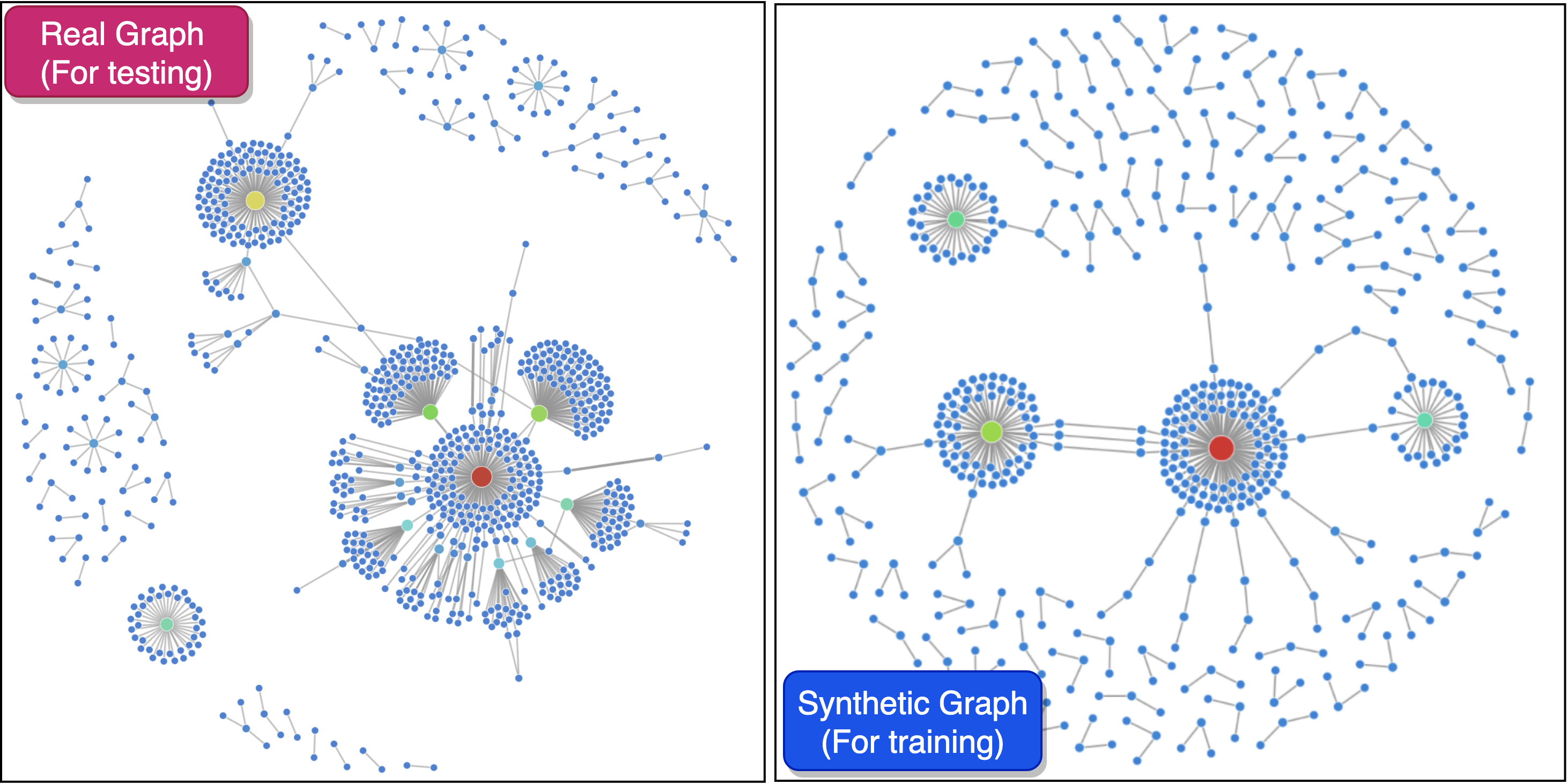}
  \caption{Examples of a real (Left) and synthetic (Right) news propagation networks on Twitter with a similar star-like shape structure. 
%   Directional arrows are omitted for the sake of simplicity.
  } 
  \label{fig:realsynthetic}
  \vspace{-15pt}
\end{figure}

\subsection{Parameterization}

A policy network $\pi_1$ is a Multi-Layer Perceptron (MLP) followed by a softmax function that projects $s^I$ to a probability distribution of 4 possible activities.
%tweet, retweet, reply, and mention. 
We can then sample $a^I$ from such a distribution. A policy network $\pi_2$ utilizes Convolutional Neural Network~\cite{kalchbrenner2014convolutional} (CNN) to efficiently extract useful \textit{spatial} features from the stack of representation vectors of all vertex $u{\in} V$ calculated by \textit{node2vec}($G$) (Sec. \ref{sec:MDP}), and MLP to extract features from the rest of the components of $s^{II}$. The resulted vectors are then concatenated as the final feature vector. Instead of directly projecting this feature on the original action space of $a^{II}$ using an MLP, we adopt the parametric-action technique~\cite{gauci2018horizon,openai2019dota} with invalid actions at each timestep $t$--i.e., already \textit{chosen} node, being masked.

\subsection{Learning Paradigm}

\textbf{Learning algorithm.} We train $\pi_1, \pi_2$ using the \textit{actor-critic} Proximal Policy Optimization (PPO) algorithm~\cite{schulman2017proximal}. It has a theoretical guarantee and is known to be versatile in various scenarios~\cite{schulman2017proximal,song2021autonomous,gogineni2020torsionnet,zhang2020learning}. The actor refers to $\pi_1$ and $\pi_2$, as described above. Their critics share the same network structure but output a single scalar value as the estimated accumulated reward at $t$.
\\

\noindent \textbf{Learning on synthetic and evaluating on real networks. }
We evaluate our method on real world data. To make our RL model generalize well on unseen \textit{real} networks (Figure \ref{fig:realsynthetic}, Left) with different possible configurations of $G{=}(V,E)$, it is important to train our model on a sufficient number of diverse scenarios--i.e., training graphs. However, collecting such a train dataset often requires much time and efforts. Hence, we propose to train our model on synthetic graphs, which can be efficiently generated on the fly during the training~\cite{kamarthi2019influence}. To avoid distribution shifts between train and test graphs, we first collect a \textit{seed dataset} of several news propagation networks and use their statistical properties ($p_{intra}, p_{inter}$) to \textit{spontaneously} generate a \textit{synthetic} graph (Figure \ref{fig:realsynthetic}, Right) for each training iteration. We describe this in detail in Section~\ref{sec:experiment}.

\section{Experiment}\label{sec:experiment}
\subsection{Set-Up}

\textbf{Datasets.} We collected a total of top-100 trending articles on Twitter from January 2021 to April 2021 and their corresponding propagation networks with a maximum of 1.5K nodes using the public Hoaxy API\footnote{\url{https://rapidapi.com/truthy/api/hoaxy}}. All the downloaded data is free from user-identifiable information. The majority of these articles are relevant to the events surrounding the 2020 U.S. presidential election and the COVID-19 pandemic. We also share the same observation with previous literature~\cite{kamarthi2019influence,sadikov2011correcting} such that retweet networks tend to have star-like shapes. These networks have a high $p_{intra}$ and a low $p_{inter}$ value, suggesting multiple separate star-shape communities with few connections among them. Therefore, viral news usually originates from a few very influential actors in social networks and quickly propagates to their followers.
\\

 \begin{figure}[t!]
  \centering
  \includegraphics[width=0.45\textwidth]{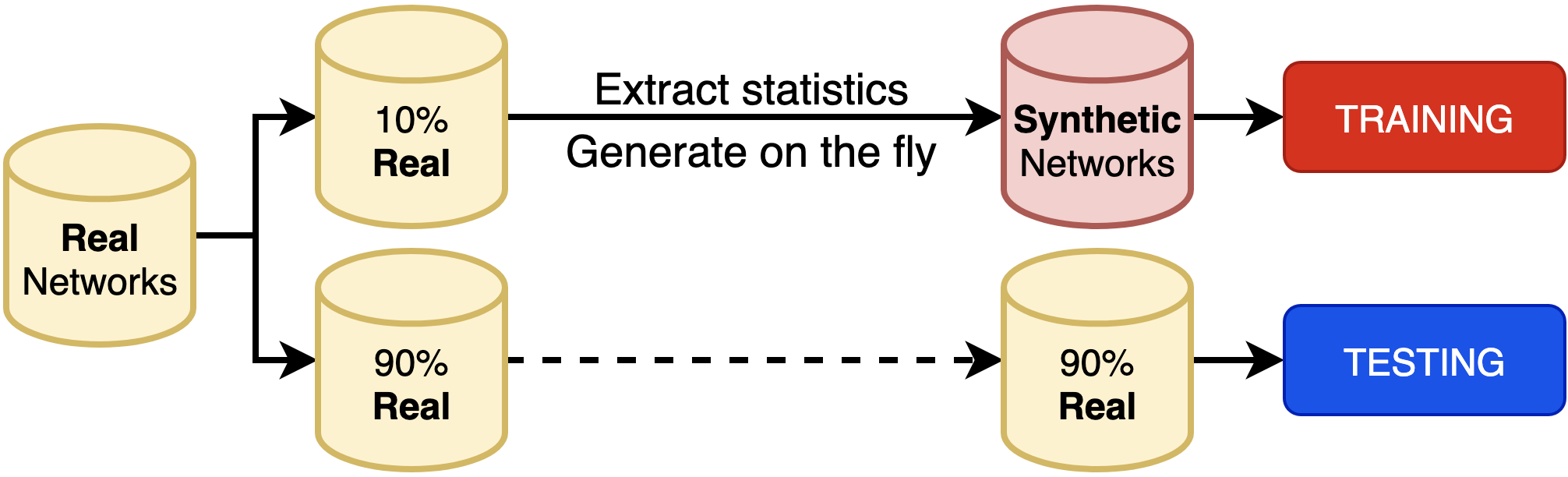}
  \caption{We generate synthetic networks that ensemble real networks' structures on the fly to train {\mymethod} and test it with real networks.} 
  \label{fig:datasplit}
  \vspace{-15pt}
\end{figure}

\begin{figure*}[t!]
  \centering
  \includegraphics[width=0.95\textwidth]{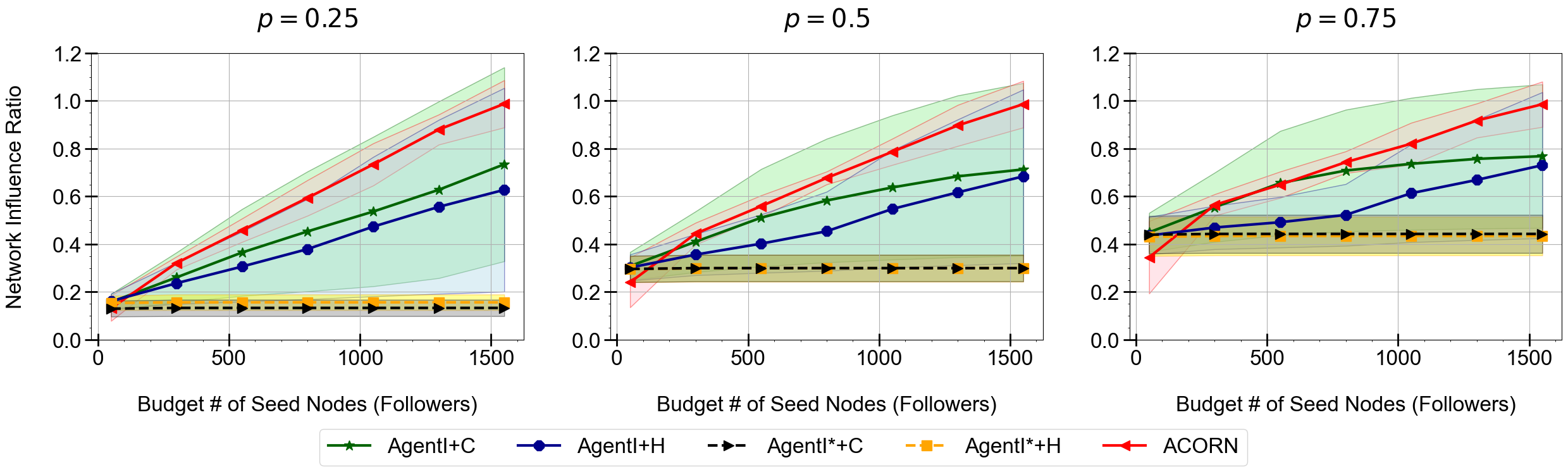}
%   \vspace{-15pt}
  \caption{Performance comparison of a single socialbot \textit{under bot detection constraint}.
%   \lee{present 5 methods in consistent order throughout: AgentI+C, AgentI+H, AgentI*+C, AgentI*+H, ACORN}
  }
  \label{fig:singlebot}
\end{figure*}

\noindent \textbf{Training and Testing Set.} Figure \ref{fig:datasplit} illustrates how to utilize synthetic data during training. Since we observe that our framework generalizes better when trained with more complex graphs--i.e., more edges with high intra-community ($p_{intra}$) and inter-community ($p_{inter}$) edge probabilities, We first selected 10\% of the collected \textit{real} networks with the highest $p_{intra}$ and $p_{inter}$ as initial \textit{seed graphs}--e.g., Figure \ref{fig:realsynthetic}, Left, to generate the training set and use the rest as the test set. Then, during training, we used the average statistics ($p_{intra}, p_{inter}$, \# of communities and their sizes) of the \textit{seed graphs} to generate a \textit{stochastic, synthetic} graph for each training episode of a maximum $T$ timesteps--e.g., Figure \ref{fig:realsynthetic}, Right. These two statistics are selected because they well capture the star-like shapes of a typical retweet network. Since the real activation probabilities $p$ of the collected networks are unknown, we found that using a fixed high $p$ value during training achieves the best results. We then reported the averaged results across 5 different random seeds on the remaining 90 real test networks with varied $p$ values and on a much longer horizon than $T$. Note that this number of testing networks is  much larger and more extensive than those of previous studies \cite{kamarthi2019influence,wen2016online,kempe2003maximizing}.
\\

\noindent \textbf{Baselines.} Since there are no previous works that address the ASL problem, we combined different approximation and heuristic approaches for the IM task with the socialbot detector evasion feature that is provided by \textit{learned} \textsc{AgentI} as baselines:
% \vspace{-5pt}
\begin{itemize}[leftmargin=\dimexpr\parindent+0.1\labelwidth\relax]
\item  \textsc{\textbf{AgentI+C}}. This baseline extends the \textit{Cost Effective Lazy Forward (CELF)}~\cite{leskovec2007cost} and exploits the submodularity of the spread function $\pmb{\sigma}(\cdot)$ to become the first substantial improvement over the traditional \textsc{Greedy} method~\cite{kempe2003maximizing} in terms of computational complexity. IM is a standard baseline in influence maximization literature.
\item \textsc{\textbf{AgentI+H}}. Since $G$ consists of several star-like communities, we also used a \textit{heuristic approach} \textsc{Degree}~\cite{kempe2003maximizing,chen2009efficient} that always selects the node with the largest out-degree that is available--i.e., user with the largest \# of followers.
\item \textsc{\textbf{AgentI*+C}} and \textsc{\textbf{AgentI*+H}} train the first-level agent \textit{independently} from the second-level agent and combined it with CELF or the heuristic approach \textsc{Degree}, respectively. These are introduced to examine the dependency between the trained \textsc{AgentI} and \textsc{AgentII}
\end{itemize}

\noindent Since the \textsc{Greedy} approach does not scale well with a large number of seeds, however, we excluded it from our experiments.
\\

\noindent \textbf{Models and Configurations.} We used a fixed hyper-parameter setting. During training, we set $K{\leftarrow}20, Q{\leftarrow}3, T{\leftarrow}60, p{\leftarrow}0.8$, and $\gamma_{\mathrm{step}},\gamma_{\mathrm{delayed}}{\leftarrow}0.99$. We refer the readers to the appendix for detailed configurations for RL agents.  We ran all experiments on the machines with Ubuntu OS (v18.04), 20-Core Intel(R) Xeon(R) Silver 4114 CPU @ 2.20GHz, 93GB of RAM and a Titan Xp GPU 16GB. All implementations are written in Python (v3.8) with Pytorch (v1.5.1).

\begin{figure*}[t]
  \centering
  \includegraphics[width=0.96\textwidth]{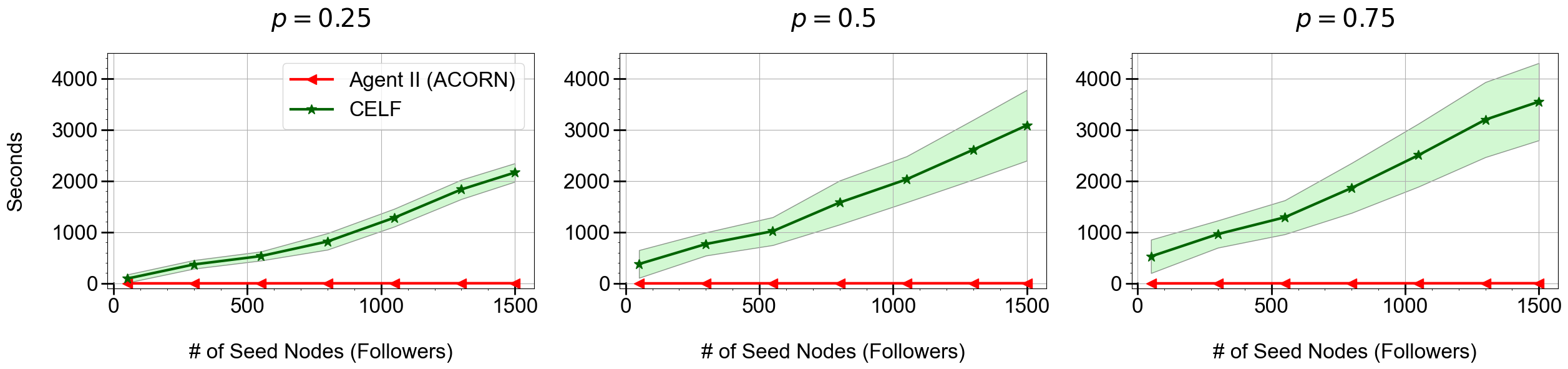}
%   \vspace{-15pt}
  \caption{Empirical comparison of running time between  \textsc{CELF} and \textsc{AgentII} ({\mymethod}).}
  \label{fig:runningtime}
\end{figure*}

\begin{figure*}[t]
  \centering
  \includegraphics[width=0.96\textwidth]{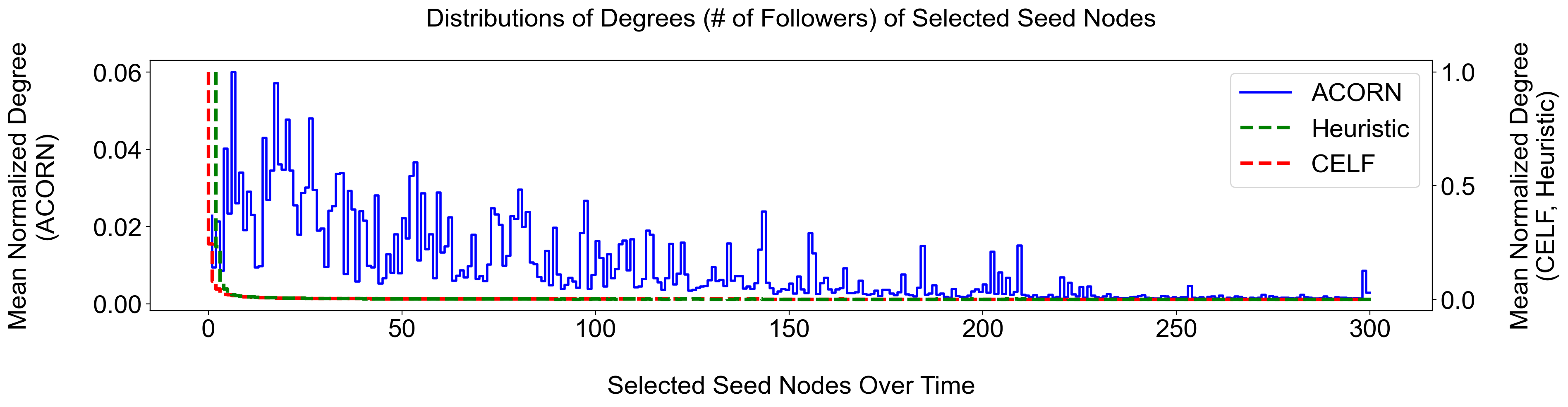} 
  \vspace{-15pt}
  \caption{Insights on the learned policies.}
  \label{fig:insights}
\end{figure*}

\subsection{Main Results}

\textbf{Network Influence Ratio.} Figure \ref{fig:singlebot} shows the network influence ratio--i.e., network influence over total number of users, \textit{under a bot detection environment} given different number of budget seeds $|S|$ and $p$ values:% $\pmb{\sigma}(\mathcal{S},G)/|V|{\leq}1.0$.
\begin{equation}
    \pmb{\sigma}(\mathcal{S},G)/|V|{\leq}1.0
\end{equation}
% ($\pmb{\sigma}(\mathcal{S},G)/|V|{\leq}1.0$).
A high network influence ratio requires both (i) efficient followship selection and (ii) efficient detection evasion strategy. Overall, \textsc{\mymethod} outperforms all baselines with different news virality ($p$ values). However, \textsc{\mymethod} underperforms when $|S|$ is low--e.g., $|S|{=}50$ in Figure \ref{fig:singlebot}. This is because \textsc{AgentII} learns not to connect with the most influential nodes early in the process. This can help prevent disrupting the sequence $\mathcal{A}$ and lead to early detection, especially when it gets closer to the next prediction interval of $\mathcal{F}$. 

The larger the $p$ value, the further--i.e., more hoops, a news can propagate through $G$. Hence, as $p$ increases--i.e., the more viral a piece of news, utilizing the network structure to make new connections is crucial and more effective than simply selecting the most influential users. This is reflected in the inferior performance of \textsc{AgentI+H} when compared with \textsc{AgentI+C}, {\mymethod} in Figure \ref{fig:singlebot}, $p{=}0.75$. This means that {\mymethod} is able to utilize the network structured capture by \textit{node2vec} and postpone short-term incentives--i.e., makes friends with influential users, for the sake of long-term rewards. Overall, \textsc{Acorn} also behaves more predictably than baselines in terms of the influence ratio's deviation across several runs.
\\

\renewcommand{\tabcolsep}{1.5pt}
\begin{table}[t]
    \centering
    \footnotesize
    \caption{Total survival timesteps v.s. network influence ratio after reaching $|S|{=}{|V|}$}
    \begin{tabular}{lcccccc}
        \toprule
        \multicolumn{1}{c}{} & \multicolumn{2}{c}{$p=0.25$} & \multicolumn{2}{c}{$p=0.50$} & \multicolumn{2}{c}{$p=0.75$} \\
        \cmidrule(lr){2-3}\cmidrule(lr){4-5}\cmidrule(lr){6-7}
        {} & \%$\uparrow$ & Steps$\uparrow$ & \%$\uparrow$ & Steps$\uparrow$ & \%$\uparrow$ & Steps$\uparrow$ \\ 
        \midrule
        % Agent I$^*$+H & 0.16 {$\pm$} 0.03 & 101 {$\pm$} 80 & 0.30 {$\pm$} 0.06 & 107 {$\pm$} 91 & 0.43 {$\pm$} 0.1 & 110 {$\pm$} 87\\
        \textsc{AgentI+H} & 0.63 {$\pm$} 0.43 & 1.2K {$\pm$} 1K & 0.68 {$\pm$} 0.36 & 1.2K {$\pm$} 1K & 0.73 {$\pm$} 0.31 & 1.2K {$\pm$} 1K\\
        % \cmidrule(lr){1-7}
        \textsc{AgentI+C} & 0.73$\pm$0.41 & 1.5K$\pm$968 & 0.71$\pm$0.36 & 1.3K$\pm$1K & 0.77$\pm$0.30 & 1.3K$\pm$1.1K\\
        \textsc{\mymethod} & \textbf{0.99} {$\pm$} \textbf{0.10} & \textbf{2.1K} {$\pm$} \textbf{254} & \textbf{0.99} {$\pm$} \textbf{0.10} & \textbf{2.0K} {$\pm$} \textbf{276} & \textbf{0.99} {$\pm$} \textbf{0.10} & \textbf{2.0K} {$\pm$} \textbf{305}\\
        \bottomrule
        % \multicolumn{7}{l}{{H}: \textsc{HEUR}, {C}: \textsc{CELF}}
    \end{tabular}
    \label{tab:survival}
    \vspace{-10pt}
\end{table}

\noindent  \textbf{Survival Timesteps.} We then evaluated if a trained socialbot can survive even after collecting all followers. Table \ref{tab:survival} shows that while we train a socialbot with a finite horizon $T{=}60$, it can live on the network for a much longer period during testing. However, other baselines were detected very early. Since only three out of four activities--i.e., tweet, retweet, reply, and mention, allow to collect new followers, it is natural that socialbots need to survive much longer than $|V|$ steps--e.g., around 2.0K in Table \ref{tab:survival}, to accumulate all followers. This corresponds to 98\%, 64\%, and 56\% of socialbots surviving--i.e., not detected, after reaching $|S|{=}V$ for \textsc{Acorn}, \textsc{AgentI+C} and \textsc{AgentI+H}, respectively. Our trained socialbot can also sustain much longer if we keep it going during testing, even with different detection intervals $K{>}20$. This implies that \textsc{AgentI} can generalize its adversarial activities against $\mathcal{F}$ toward unseen real-life scenarios.
\\

\noindent \textbf{Dependency between  RL Agents.} The above results also demonstrate the effects of co-training \textsc{AgentI} and \textsc{AgentII}. First, the heuristic and CELF method when paired with the learned \textsc{AgentI} (blue \& green lines, Figure \ref{fig:singlebot}) performs much better than when paired with an \textit{independently} trained (without \textsc{AgentII}) \textsc{AgentI} (yellow \& black lines, Figure \ref{fig:singlebot}). This shows that \textsc{AgentI}, when trained with \textsc{AgentII}, becomes more versatile and can help a socialbot survive a much longer period of time, especially even when the socialbot only uses a heuristic node selection. However, \textsc{AgentI} performs the best when paired with \textsc{AgentII}. This shows that two RL agents successfully learn to collaborate, not only to evade the socialbot detection but also to effectively maximize its network influence. This further reflects the co-dependency between the roles of $\mathcal{A}$ and $\mathcal{S}$ as analyzed in Sec. \ref{sec:environment}.
\\
\begin{figure*}[t]
  \centering
  \includegraphics[width=0.96\textwidth]{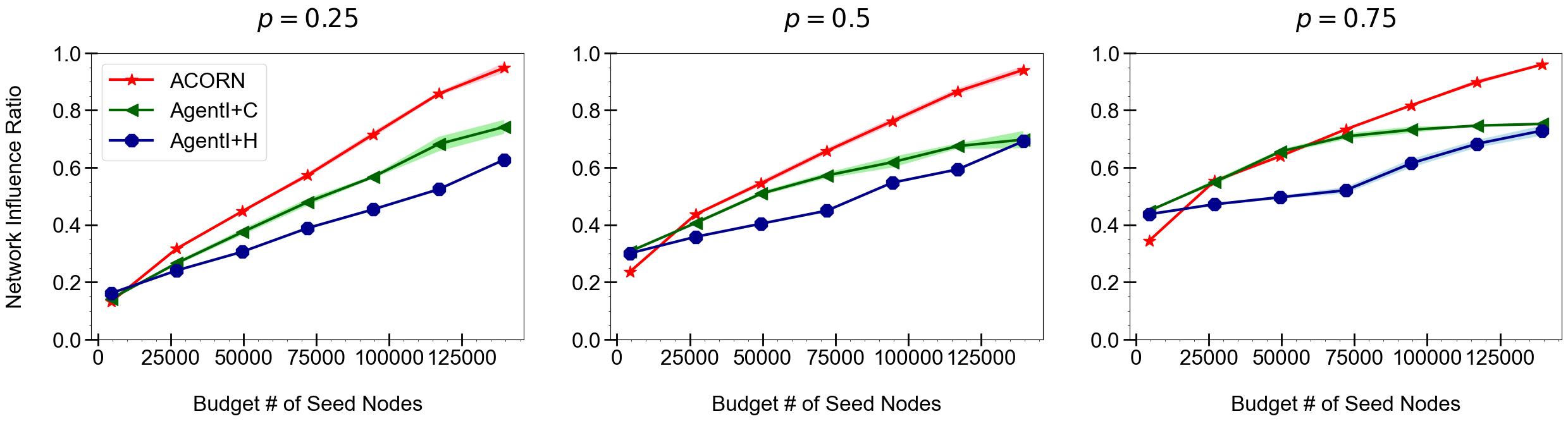}
%   \vspace{-15pt}
  \caption{Performance of multiple socialbots \textit{under bot detection constraint} on a large network.}
  \label{fig:multibots}
  \vspace{-5pt}
\end{figure*}

\noindent \textbf{Computational Analysis.} We compared the computational complexity of \textsc{AgentII} specifically with the \textsc{CELF} algorithm during inference. Even though \textsc{CELF} significantly improves from the traditional \textsc{Greedy}~\cite{kempe2003maximizing} IM algorithm with the computational complexity of $\mathcal{O}(|S||V|m)$~\cite{tang2014influence} (assuming each call of $\pmb{\sigma}$ takes $\mathcal{O}(m)$ and only one round of Monte Carlo simulation is needed), its computation greatly depends on  $\pmb{\sigma}(\cdot)$, the size of the graph and becomes only computationally practical when $|S|$ is small. This is also similar to other traditional IM algorithms such as \textsc{CELF++}~\cite{goyal2011celf++}, TIM~\cite{tang2014influence}, and ASIM~\cite{galhotra2015asim}. To illustrate, \textsc{CELF} takes much more time to compute as $|S|$ increases especially with large $p$--i.e., more nodes need to be reached when computing $\pmb{\sigma}(\cdot)$ (Figure \ref{fig:runningtime}). However, with the $\mathcal{O}(1)$ complexity of the forward pass through $\pi_2$, \textsc{AgentII} is able to scale linearly $\mathcal{O}(|S|)$ regardless of the network structure and the virality of the news during inference. Even though our framework requires to  calculate the graph representation using \textit{node2vec}, it is specifically designed to be scalable to be able to process large graphs~\cite{rossi2018deep} and we only need to run it \textit{once}.  
\\

\noindent \textbf{Insights on the Learned Policies.} We summarized the node selection strategies of all methods in Figure \ref{fig:insights}. We observed that both heuristic and \textsc{CELF} selects very influential nodes with many followers (high out-degrees) very early. Alternatively, \textsc{AgentII} acquires an array of normal users (low out-degrees) before connecting with influential ones. This results in early detection and removal of the baselines and sustainable survival of our approach. This shows that \textsc{AgentII} can learn to cope with the relationship constraint (Eqn. (\ref{eqn:dependency})) between $\mathcal{A}$ and $\mathcal{S}$ imposed by the environment. Moreover, the degrees of selected users by {\mymethod} has a \textit{right} long-tail distribution, which means that \textsc{\mymethod} overall still tries to maximize its network influence early in the process.

\subsection{Multiple Socialbots Results}

We have evaluated our approach on different real-life news propagation graphs. These networks can be considered as sub-graphs of a much larger social network. In practice, different sub-graphs can represent different communities of special interests--e.g., politics, COVID-19 news, or different characteristics--e.g., political orientation. Since socialbots usually target to influence a specific group of users--e.g., anti-vaxxer, it is practical to deploy several bots working in tandem on different sub-graphs. To evaluate this scenario, we aggregated all 90 test sub-graphs into a large network of 135K nodes and used each learned socialbot for each sub-graph. Figure \ref{fig:multibots} shows that {\mymethod} still outperforms other baselines especially later in the time horizon. Moreover, {\mymethod} can efficiently scale to a real-life setting thanks to its linear running time and highly parallel architecture.

\section{Discussion and Limitation}
% \subsection{Benefits of Multi-Agent HRL Framework}
% \subsection{Potential Applications and Misuse}
Our contribution goes beyond our demonstration such that one can train adversarial socialbots to effectively navigate real-life networks using an HRL framework. We will also publish a multi-agent RL environment for the ASL task under the \textit{gym} library~\cite{gym}. This environment will facilitate researchers to test different RL agents, examine and evaluate assumptions regarding the behaviors of socialbots, bot detection models, and the underlying influence diffusion models on synthetic and real-life news propagation networks. It remains a possibility that our proposed framework could be deliberately exploited to train and deploy socialbots to spread low-credibility content on social networks without being detected. To reduce any potential misuse of our work, we have also refrained from evaluating our framework with an actual socialbot detector API such as \textit{Botometer}~\footnote{\url{https://botometer.osome.iu.edu/}}. However, ultimately, such misuse can occur (as much as the misuse of the latest AI techniques such as GAN or GPT  is unavoidable). Yet, we firmly believe that the benefits of our framework in demonstrating the possibility of adversarial nature of socialbots, and enabling researchers to understand and develop better socialbot detection models far outweigh the possibility of misuse for developing ``smarter" socialbots. In fact, by learning and simulating various adversarial behaviors of socialbots, we can now analyze the weakness of the current detectors. Moreover, we can also incorporate these adversarial behaviors to advance the development of novel bot detection models in a \textit{proactive} manner~\cite{cresci2020decade}. Time-wise, this gives us a great advantage over the traditional \textit{reactive} flow of developing socialbot detectors where researchers and network administrators are always one step behind the malicious bots developers~\cite{cresci2020decade}.

% \subsection{Limitation}
One limitation of our current approach is that we only considered statistical features of a bot detector that are relevant to four activities--i.e., tweet, retweet, reply, and mention (Table \ref{tab:features}). While these features help achieve 90\%  of detection accuracy in F1 score on a real-life dataset, we hope to lay the foundation for further works to consider more complex network and content-based features~\cite{efthimion2018supervised,botwpi,yang2020scalable,mazza2019rtbust}. 

\section{Conclusion and Future Work}

This paper proposes a novel {\em adversarial socialbot learning} (ASL) problem where a socialbot needs to simultaneously maximize its influence on social networks and minimize the detectability of a strong black-box bot detector. We carefully designed and formulated this task as a cooperative game between two functional {\em hierarchical reinforcement learning} agents with a global reward. We demonstrated that the learned socialbots can sustain their presence on unseen real-life networks over a long period while outperforming other baselines in terms of network influence. During inference, the complexity of our approach also scales linearly with the number of followers and is independent of a network's structures and the virality of the news. Our research is also the first step towards developing more complex adversarial socialbot learning settings where multiple socialbots can work together to obtain a common goal~\cite{cresci2020decade}. By simulating the learning of these socialbots under various realistic assumptions, we also hope to analyze their adversarial behaviors to develop effective detection models against more advanced socialbots in the future. \footnote{{The work was in part supported by NSF awards
\#1820609, \#1940076, and \#1909702}}

\pagebreak

\bibliographystyle{ACM-Reference-Format}
\bibliography{sample-base}

\appendix

\end{document}